\documentclass[%
 aip,
 jmp,%
 amsmath,amssymb,
 reprint,%
]{revtex4-1}

\usepackage{graphicx}
\usepackage{dcolumn}
\usepackage{bm}
\usepackage[mathlines]{lineno}

\begin{document}

\preprint{AIP/123-QED}

\title{Vibrational Spectroscopy of Methyl benzoate}

\author{Kiran Sankar Maiti}
 \email{kiran.maiti@ch.tum.de.}
 \affiliation{Department of Chemistry and Molecular Biology, University of Gothenburg, Box 462, SE-40530 Gothenburg, Sweden.}
\author{Christoph Scheurer}%
\affiliation{ 
Department of Chemistry, 
Technical University of Munich, D-85747 Garching, Germany. 
}%

\date{\today}

\begin{abstract}
Methyl benzoate (MB) is studied as a model compound for the development of 
new IR pulse schemes with possible applicability to biomolecules. Anharmonic 
vibrational modes of MB are calculated on different level (MP2, SCS, CCSD(T)
with varying basis sets) ab-initio PESs using the vibrational self-consistent 
field (VSCF) method and its correlation corrected extensions. Dual level 
schemes, combining different quantum chemical methods for diagonal and coupling 
potentials, are systematically studied and applied successfully to reduce 
the computational cost. Isotopic substitution of $\beta$-hydrogen by deuterium 
is studied to obtain a better understanding of the molecular vibrational 
coupling topology.

\end{abstract}

\maketitle

\begin{quotation}
\end{quotation}

\section{Introduction}

Vibrational spectroscopy is among the foremost experimental tools in the
exploration of molecular potential-energy surfaces (PES). Its application 
to biological systems has so far been severely handicapped, both by
experimental difficulties and by the unavailability of adequate computational
tools for quantitative interpretation. Recent success in the experimental
realization of coherent multidimensional infra-red (IR) spectroscopy provides
a  new powerful tool to study structure and dynamics of biomolecules with a
temporal resolution down to the sub-picosecond regime\cite{hamm1998:6123, 
borm2000:41,jona2003:425, brix2005:625,zhen2006:1951,tokm2007:54}. 
Multidimensional IR spectroscopy has the potential to disentangle the 
congested vibrational spectra of biomolecules to some extent similar to 
multidimensional NMR\cite{sche2001:3114, sche2001:4989, sche2002:6803} but 
with significantly higher temporal resolution. In nonlinear multidimensional 
spectra the structural and dynamical information is typically available in 
terms of diagonal and cross-peak shapes, locations and intensities and their 
respective temporal evolution\cite{sche2007:503}. The interpretation of this 
data in terms of a dynamical model of the biomolecule under investigation 
requires extensive theoretical modeling.

The calculation of vibrational spectra within the harmonic approximation is 
very useful, but often has limited significance since many biologically 
relevant molecules are ``floppy'' and subject to strong anharmonic
effects\cite{wang2006:3798}. Anharmonic effects are even much larger in 
weakly bound molecular complexes, e.g. hydrogen-bonded complexes with 
surrounding water\cite{gerb2002:142}. Also, frequently one is interested in the
regions of the PES far away from the equilibrium configuration, where the 
harmonic approximation is even less applicable. The main problem of anharmonic 
spectroscopic calculations is that different vibrational modes are not 
mutually separable like in the harmonic approximation\cite{boun2006:87,
boun2008:194}. Therefore one has to face  the task of
calculating  wavefunctions and energy levels for systems of many coupled
degrees of freedom. Several attempts have been made to overcome this
problem. Among others, the discrete variable representation
(DVR)\cite{baci1986:3606,hend1993:7191, wrig1999:902}, diffusion
quantum Monte Carlo (DQMC)\cite{ande1975:1499, buch1992:726, barn1993:9730},  
and vibrational self-consistent-force field
(VSCF)\cite{bowm1978:608, gerb1979:195, gerb1988:97}, methods proved their
applicability to study anharmonic effects in systems with varying sizes. The
VSCF method is most successful among them to effectively  handle large
molecular systems.

IR absorption spectra of peptides and proteins are dominated by vibrational 
bands that can be described approximately in terms of oscillators localized 
in each repetitive unit and their mutual couplings. The most extensively 
studied bands are amide-A and amide-B in the region 3000--3500\,cm$^{-1}$ and 
amide-I and amide-II between 1500-1700\,cm$^{-1}$, which are spectrally well 
separated from the remaining spectrum and exhibit a strong dependence on the 
structural motifs present in the investigated biomolecules\cite{bart2002:369,
mait2012:16294}. The amide-I vibrational mode, which  involves
mainly the C=O stretch coordinate, has experimentally been the most important
mode due to its large transition-dipole moment and because it appears to be
mostly decoupled from the remaining vibrational modes in proteins. A detail
understanding of these modes is then necessary to understand the structure
and dynamics of the protein and peptide. Due to the complex structure
of proteins a detailed theoretical understanding of these modes is
complicated. A simple small  molecule is then necessary for which
these exemplay modes and their mutual couplings can be studied in more detail.

\begin{figure}[!ht]
\begin{center}
\includegraphics[width=0.8\linewidth,angle=0.0,clip]{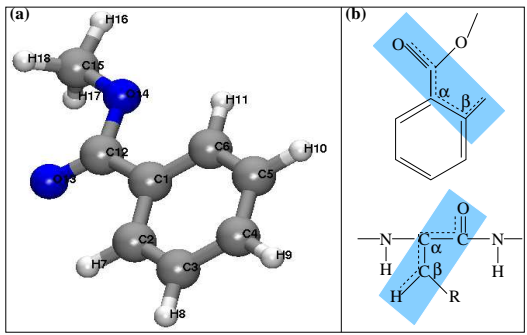}
\caption{(a) Structure of Methyl benzoate, indicating the numbering scheme 
used in this work. (b) Structural similarity of protein backbone and Methyl 
benzoate.}\label{fig:meth-benzoate-no}
\end{center}
\end{figure}

In this work Methyl benzoate was chosen as a model for potential vibrational 
couplings between the carbonyl group and a sidechain C-H moiety, which would 
allow for a determination of backbone-sidechain dynamics in peptides and 
proteins. Methyl benzoate has an idealized planar configuration, except for 
the two hydrogen atoms at the methyl
group which are symmetrically out-of-plane with respect to the rest of the
molecule (see Fig.~\ref{fig:meth-benzoate-no}).  The methyl carboxylate group 
is then co-planar with the phenyl ring. The C=O double bond in the
carboxylic ester group essentially constitutes a local oscillator similar to 
the amide-I band in proteins and provides a convenient mode for model studies.  
A potentially interesting coupling in proteins is the coupling between
the amide-I and a $\beta$-hydrogen located in the sidechain. Methyl benzoate
also provides a similar structural motif  where the ortho hydrogen in the 
phenyl ring provides the counter-part of a $\beta$-hydrogen in protein 
sidechains. The structural similarities of these two motifs in Methyl benzoate 
and proteins are depicted in Fig.~\ref{fig:meth-benzoate-no}b.

Isotopic substitution is a valuable tool in the identification of molecular 
structure and dynamics. The C--H or C--D stretch vibrations are particularly 
important structural probes, because they are very localized, specific, and 
abundant. Especially the C--D stretching frequency is an excellent structural 
probe since it is usually spectrally isolated ($\sim$2200\,cm$^{-1}$) even in 
the spectrum of large proteins. Substitution of $\beta$-hydrogen by deuterium 
provides a probe to obtain a better understanding of the structure and 
dynamics of protein sidechains with respect to the backbone. It is thus an 
interesting task to devise new experiments which probe this relative geometry. 
Methyl benzoate is an ideal candidate model which, once its vibrational 
hamiltonian is fully understood, will serve to develop new pulse sequences.

\section{Computational methods}
\subsection{Quantum Chemistry}
Geometry optimization of Methyl benzoate was performed using  second order
M{\o}ller-Plesset\cite{moll1934:618} (MP2) perturbation
theory\cite{gord1988:503, fris1990:275} and employing the augmented
correlation-consistent polarized-valence-triple zeta (aug-cc-pVTZ) basis
set. Harmonic normal-mode analysis was performed with the density fitting MP2
(DF-MP2) method using an aug-cc-pVTZ regular basis and cc-pVTZ fitting basis
sets. All calculations were performed with the MOLPRO quantum chemistry
program\cite{MOLPRO_brief}.

The choice of the computational method and the basis set are not arbitrary.
A standard MP2 with aug-cc-pVTZ basis set level of computation is generally a
reliable method to generate an anharmonic PES. For a system like Methyl 
benzoate it is not
suitable due to the size of the molecule and the resulting high computational
cost. To find a suitable method a comparison of computations employing
different basis sets has been performed and is presented in
Table~\ref{tab:df-mp2-comp}. The standard MP2 calculation employing an
aug-cc-pVTZ basis set is given in the first row and provides the 
reference. It is clear that the standard MP2 method with aug-cc-pVTZ basis is 
beyond our scope to study all 48 normal modes for Methyl benzoate, as it 
takes almost one day and a very large memory space for a
single-point energy calculation. If no specially optimized fitting basis sets
are available for a certain regular AO basis set it is common practice for the
DF-MP2 computations to use a fitting basis set one order higher than the
regular one. The DF-MP2 with  aug-cc-pVTZ regular basis and cc-pVQZ
fitting basis speeds up the calculation dramatically (see
Table~\ref{tab:df-mp2-comp}) without sacrificing quality compared to the
standard  MP2/aug-cc-pVTZ results.  
\begin{table}[!h]
\caption{Single point energy, calculation time and the required memory for
energy calculation of Methyl benzoate at equilibrium geometry with different 
basis sets. } 
\label{tab:df-mp2-comp}
\begin{ruledtabular}
\begin{tabular}{ccrrrrr}
\multicolumn{2}{c}{Basis} & \multicolumn{3}{c}{Energy in E$_h$} & CPU time & Memory\\
\cline{1-5}
mp2fit & jkfit & SCF & MP2 & Total & in Sec. & in MB\\
\hline
     &     & -457.5104 & -1.7604 & -459.2708 & 85363 & 6799.36\\
avtz & vtz & -457.5101 & -1.7601 & -459.2702 &  2822 & 270.39\\
avtz & vqz & -457.5103 & -1.7603 & -459.2706 &  5829 & 361.65\\
avqz & vqz & -457.5374 & -1.8608 & -459.3981 & 17132 & 3194.88 \\
avqz & v5z & -457.5374 & -1.8608 & -459.3982 & 29014 & 3194.88 \\
\end{tabular}
\end{ruledtabular}
\end{table}
A single-point energy calculation with this method takes only one and a half 
hours and requires almost 200 times less memory than memory require for regular
MP2 calculation with aug-cc-pVTZ basis set. Whereas the DF-MP2 method
employing aug-cc-pVTZ regular and cc-pVTZ fitting basis sets gives almost
identical energy values in half the time and is a reasonable choice for our 
purpose. The anharmonic pair couplings are calculated with the DF-MP2 level of 
theory employing the same cc-pVDZ basis for regular and fitting basis set.  
For some of the most problematic
modes the diagonal potentials are calculated with the local density fitting
coupled cluster singles and doubles and  perturbative triple correction
DF-L-CCSD(T)\cite{schu2000:370, schu2002:3941, schu2003:3349}  method with the
cc-pVTZ basis set in a dual-level scheme. All computations were additionally 
performed using the density fitting spin-component scaled
(DF-SCS) MP2 method\cite{grim2003:9095} employing different basis sets.

\subsection{Grid for VSCF}
The choice of an appropriate grid size is crucial for anharmonic frequency
calculations for both diagonal and pair potentials.
The VSCF PES can be expressed in terms of a hierarchical 
expansion\cite{cart1997:1179} 
\begin{eqnarray}V(q_1, \cdots, q_N)&=&\sum_j^N V_j^{(1)}(q_j)+ \sum_{i<j}
  V_{i,j}^{(2)}(q_i,q_j)\nonumber \\
  &+&\sum_{i<j<k}V_{i,j,k}^{(3)}(q_i,q_j,q_k)+\cdots  \nonumber\\ 
&+&\sum_{i<j\cdots<r<s}V_{i,j,\cdots,r,s}^{(n)}(q_i,q_j,\cdots,q_r,q_s)+\cdots.
\label{eq:vscf-13}\end{eqnarray}
where $V_j^{(1)}(q_j)$ is the diagonal potential, $V_{i,j}^{(2)}(q_i,q_j)$ is
the pairwise potential, $V_{i,j,k}^{(3)}(q_i,q_j,q_k)$ is the triple coupling
and so on. 

\begin{figure}[!h]
\begin{center} 
\includegraphics[width=0.8\linewidth,angle=0.0,clip]{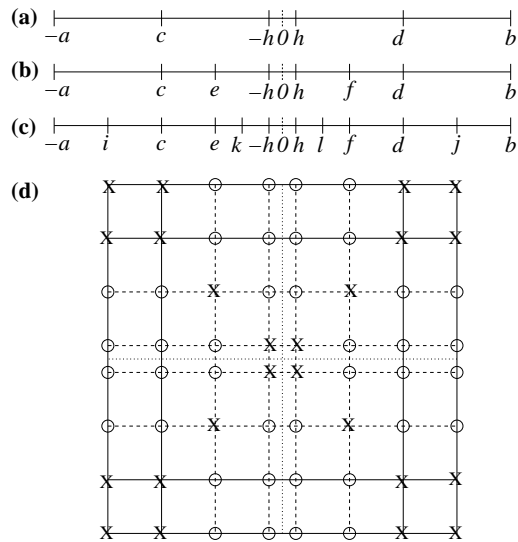}
\caption{Diagonal grid with different grid size; (a) 6 points grid, (b) 8 
points grid, (c) 12 points grid. (d) Irregular spaced 2-dimensional grid. Energy points are calculated at 
the grids, marked with crosses. The grids marked with  circles are filled up
by 2-dimensional IMLS interpolation.}\label{fig:1D-grid}
\end{center}
\end{figure}

The grid was constructed based on the harmonic frequency analysis
and PM3 PES cuts. The innermost two points (-h, h) are determined from the
second derivative of the PES at equilibrium corresponding to the harmonic
frequencies of the normal modes. The outer two points (-a, b) are determined
as those points on the 1D-PES cuts along the normal modes for which the PM3
energy reaches six times the harmonic frequency quantum w.r.t. $V_0$.  
The remaining points are calculated dividing the
interval with different proportions e.g., 
$c=\frac{-a+(-h)}{2}$, $d=\frac{b+h}{2}$, $e=\frac{c+(-h)}{2}$,
$f=\frac{d+h}{2}$, $i=\frac{-a+c}{2}$, $j=\frac{d+b}{2}$,
$k=\frac{e+(-h)}{2}$ and $l=\frac{f+h}{2}$. 
Diagonal potentials with 
6 points (see Fig.~\ref{fig:1D-grid}a) are insufficient for a reasonable
description of the PES in the calculation of anharmonic frequencies. In
particular, a denser set of grid points is required near the equilibrium
configuration. A comparative  study has been carried out with 8 point (see
Fig.~\ref{fig:1D-grid}b) and 12 point (see Fig.~\ref{fig:1D-grid}c) 1D
grids. It indicates that the 8 point grid PES is a reasonable choice
for anharmonic frequency calculations.

All diagonal points were first evaluated on an eight point one-dimensional 
grid as shown in Fig.~\ref{fig:1D-grid}b, then interpolated to equidistant 
16 point grids which were used in the collocation treatment. Pair potentials, 
calculated at semiempirical PM3 level, were evaluated on $8\times8$ point 
direct product grids and then interpolated to $16\times16$ point grids by 
two-dimensional cubic spline interpolation\cite{boor:pgs}. To lower the 
computational cost while maintaining high quality pair potentials, we 
performed DF-MP2/cc-pVDZ computations for selected coupling potentials on 
irregularly spaced two-dimensional grids, as shown in Fig.~\ref{fig:1D-grid}d. 
Energies are only calculated for the grid points marked with crosses. The 
undetermined points (marked with circles in Fig.~\ref{fig:1D-grid}d) are filled
by using non-uniform IMLS (interpolating moving list-squares)\cite{lanc:imls}
interpolation, which provides a potential on $8\times8$ point direct product
grids. The $8\times8$ point grid is extended to the $16\times16$ point grid 
by using 2D cubic-spline interpolation.  

\section{Results and Discussion}
\subsection{Vibrational spectrum of Methyl benzoate}
There are not many experimental results (except for the low-resolution IR
Raman spectrum of Chattopadhyay\cite{chat1968:335})   available for 
Methyl benzoate. Also not all  modes are resolved and identified
experimentally, especially for low-frequency and near-degenerate modes.
We therefore choose our best computational result, which agrees comparatively
well with the known experimental results\cite{chat1968:335, gree1977:583,
  wils1934:706, whif1956:1350}, as a reference and then discuss all other
levels of computation with respect to that. For this purpose, we choose the 
non-correlated VSCF results as a reference, where diagonal and pair potentials 
are calculated using the DF-SCS method employing a cc-pVDZ basis set. 

The entire vibrational spectrum of Methyl benzoate can be divided into three 
different frequency regions. Most interesting high frequency modes are due to
the C-H stretching  modes. Low frequency (up to $\sim$1000\,cm$^{-1}$) modes
are primarily due to bending motions involving the phenyl ring and the ester 
group. Between these two extremes there are modes  with combinations of 
bending and stretching motions and also the spectroscopically well-separated 
C=O band at about 1725\,cm$^{-1}$.

\begin{table*}
\caption{Vibrational frequencies of Methyl-Benzoate with
 different level of theories, where diagonal grids are calculated with
 the DF-MP2/AVTZ level. Mode numbers are based on normal mode frequencies. 
The RMSD calculated with respect to the anharmonic frequency calculated with 
VSCF/DF-SCS method. The modes for which diagonal anharmonicities are observed 
larger than 10\,cm$^{-1}$ are with bold numbers.}\label{tab:final-allH} 
\begin{ruledtabular}
\begin{tabular}{crrrrrrrrrl}
Mode&Harmonic & Diagonal&\multicolumn{2}{c}{VSCF}& {VC-MP2}& VSCF/ &
Experiment & \multicolumn{2}{c}{Isotopomer} & Assignment\footnotemark[1]\\
\cline{4-5}\cline{9-10}
& & &  PM3 &  MP2 & & DF-SCS   &[Ref.~\onlinecite{chat1968:335}]& D$_7$ & D$_{11}$ &\\
\hline
1  &   52  &   82 & 147 & 103 &   87 &  112 &      &  111 &  111  & Phenyl-ester opr            \\
2  &  112  &  154 & 178 & 192 &  160 &  196 &      &  197 &  198  & Phenyl opb, ester opb       \\
3  &  185  &  361 & 366 & 316 &  352 &  342 &      &  344 &  343  & CH$_3$ asb                  \\
4  &  167  &  170 & 169 & 173 &  169 &  179 &  134 &  177 &  177  & Phenyl-ester ipr            \\
5  &  208  &  217 & 236 & 239 &  234 &  242 &  218 &  243 &  242  & Phenyl-ester opb            \\
6  &  331  &  333 & 326 & 334 &  331 &  340 &      &  339 &  340  & Phenyl-Ester ipr            \\
7  &  358  &  358 & 356 & 366 &  365 &  363 &  360 &  361 &  361  & O=C-O ipb, phenyl ipb       \\
8  &  404  &  411 & 419 & 416 &  414 &  421 &      &  402 &  403  & phenyl opb                  \\
9  &  454  &  458 & 477 & 470 &  466 &  468 &      &  458 &  461  & phenyl opb, O=C-O-C opb     \\
10 &  481  &  481 & 481 & 476 &  475 &  477 &      &  474 &  471  & phenyl ipr, O=C-O-C ipr     \\
11 &  617  &  617 & 618 & 612 &  611 &  617 &  630 &  612 &  612  & phenyl ipd                  \\
12 &  678  &  679 & 685 & 676 &  673 &  681 &  674 &  678 &  680  & phenyl ipd O=C-O ipb        \\
{\bf 13} &  712  &  728 & 788 & 737 &  726 &  754 &      &  708 &  704  & phenyl opd O=C-O opb        \\
{\bf 14} &  795  &  801 & 840 & 808 &  837 &  813 &  782 &  829 &  834  & phenyl opd O=C-O opb        \\
15 &  831  &  830 & 818 & 818 &  812 &  823 &  820 &  822 &  821  & O=C-O ipb, phenyl, CH$_3$ ip\\
16 &  694  &  760 & 759 & 781 &  724 &  731 &  714 &  734 &  725  & phenyl opd                  \\
{\bf 17} &  863  &  894 &  880 & 856 &  841 &  891 &  864 & &            & phenyl opr                  \\
{\bf 18} & 1003  & 1004 & 1170 & 974 &  971 &  978 &      &  759 &  771  & phenyl ipd, ester ipd       \\
{\bf 19} & 1012  & 1012 & 1052 & 991 &  989 &  979 &      &  973 &  973  & phenyl ipd                  \\
20 & 1048  & 1049 & 1146 & 1029 & 1025 & 1035 & 1027 &      &       & phenyl ipd, P-CH sipb       \\
{\bf 21} &  977  & 1000 & 980 & 978 &  987 &  986 &  980 &  984 &  984  & phenyl opd, P-CH opb        \\
22 & 1097  & 1101 & 1162 & 1082 & 1074 & 1096 & 1097 & 1047 & 1048  & phenyl ipd, P-CH ipb        \\
{\bf 23} &  938  &  970 & 951 & 931 &  926 &  958 &  950 &  956 &  980  & phenyl opd, P-CH opb        \\
24 & 1142  & 1144 & 1220 & 1111 & 1103 & 1126 & 1128 & 1112 & 1121  & OCH$_3$ r, O=C-O ipb        \\
{\bf 25} & 1174  & 1189 & 1154 & 1153 & 1146 & 1167 & 1161 & 1171 & 1173  & pCH ipb                     \\
{\bf 26} & 1185  & 1199 & 1074 & 1170 & 1161 & 1182 &      & 1180 & 1181  & CH$_3$ sopb, C-O-C opb      \\
{\bf 27} & 1191  & 1198 & 1176 & 1164 & 1156 & 1183 & 1177 & 1149 & 1141  & pCH sipb                    \\
{\bf 28} &  958  &  994 &  908 &  968 &  982 &  979 &      &  969 &  984  & pCH opb                     \\
29 & 1221  & 1227 & 1164 & 1200 & 1194 & 1214 & 1192 & 1212 & 1213  & OCH$_3$ r, O=C s, C-O s     \\
30 & 1317  & 1321 & 1350 & 1283 & 1262 & 1303 & 1278 & 1298 & 1307  & OCO ipb pCH ipb             \\
31 & 1330  & 1334 & 1287 & 1301 & 1309 & 1320 & 1295 & 1268 & 1259  & pCH ipb                     \\
32 & 1472  & 1469 & 1318 & 1423 & 1415 & 1342 & 1310 & 1444 & 1447  & CH$_3$ asb                  \\
33 & 1479  & 1482 & 1354 & 1433 & 1411 & 1447 & 1435 & 1348 & 1345  & pCC s                       \\
34 & 1471  & 1475 & 1461 & 1436 & 1434 & 1452 & 1444 & 1445 & 1455  & pCC s                       \\
35 & 1509  & 1507 & 1320 & 1459 & 1454 & 1453 &      & 1454 & 1453  & CH$_3$ wagging              \\
{\bf 36} & 1519  & 1517 & 1322 & 1488 & 1488 & 1485 &      & 1485 & 1485  & CH$_3$ sb                   \\
37 & 1518  & 1519 & 1529 & 1478 & 1477 & 1498 & 1474 & 1480 & 1483  & pCC s, pCH ipb              \\
38 & 1634  & 1636 & 1750 & 1587 & 1581 & 1606 & 1594 & 1602 & 1600  & phenyl ipd-sy               \\
39 & 1634  & 1634 & 1747 & 1587 & 1580 & 1603 & 1585 & 1598 & 1604  & phenyl ipd-sy               \\
40 & 1766  & 1759 & 1915 & 1724 & 1724 & 1778 & 1724 & 1724 & 1724  & C=O s                       \\
{\bf 41} & 3092  & 3059 & 2909 & 2946 & 2795 & 2974 & 2998 & 2974 & 2973  & mCH ss                      \\
{\bf 42} & 3183  & 3255 & 2787 & 2582 & 2492 & 2560 & 2542 & 2561 & 2559  & mCH as                      \\
{\bf 43} & 3205  & 3208 & 2887 & 2893 & 2849 & 2855 & 2855 & 2809 & 2889  & pCH as                      \\
{\bf 44} & 3214  & 3247 & 2867 & 2872 & 2834 & 2834 & 2852 & 2171 & 2971  & pCH as/pCD$_7$ as           \\
{\bf 45} & 3217  & 3213 & 2868 & 2860 & 2694 & 2832 & 2845 & 2858 & 2832  & mCH ss                      \\
{\bf 46} & 3222  & 3233 & 2913 & 2995 & 2849 & 2963 & 2952 & 2831 & 2957  & pCH as                      \\
{\bf 47} & 3226  & 3191 & 2932 & 3014 & 3035 & 3039 & 3064 & 3017 & 3065  & pCH as                      \\
{\bf 48} & 3238  & 3161 & 2898 & 2933 & 2928 & 2985 & 3022 & 2989 & 2161  & pCH as/pCD$_{11}$ as         \\
\hline
RMSD& 148 & 151 &  78 & 24 & 45 &&\\  
\end{tabular}\\
\end{ruledtabular}
\footnotetext[1]{Symbols: R = rotation, r = rocking, ipr = in-plane rotation,
 opr = out-of-plane rotation, ipb = in-plane bending, opb = out-of-plane
 bending, ipd = in-plane deformation, opd = out-of-plane deformation, as =
 asymmetric stretch, ss = symmetric stretch, asb = asymmetric bending, sb =
 symmetric bending, sipb = symmetric in plane bending, mCH = methyl CH bond,
 pCH = phenyl CH bond, sy = symmetric.}

\end{table*}

\begin{figure}[!h]
\begin{center}
\includegraphics[width=1.0\linewidth,angle=0.0,clip]{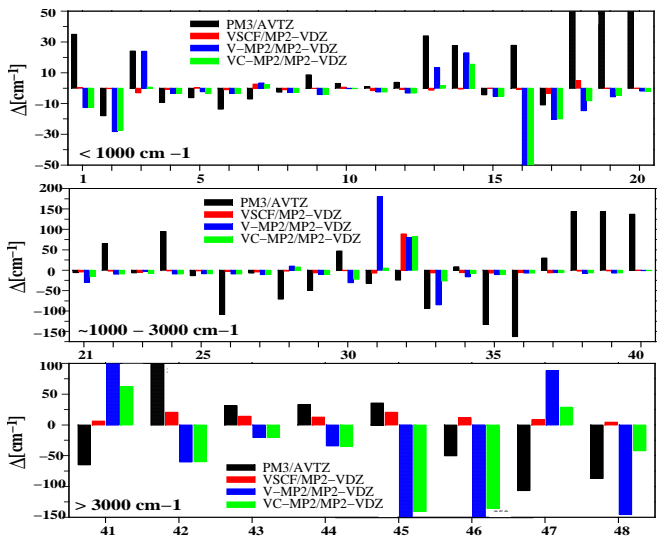}
\caption{A comparison of vibrational frequencies of Methyl benzoate calculated 
at different levels of theory w.r.t. the VSCF/DF-SCS level of calculation. The
labels indicate the level of theory was used for the 
calculation.}\label{fig:bar-p-1}
\end{center}
\end{figure}

The vibrational frequencies of all modes of Methyl benzoate calculated at 
different levels of theory are presented in Table~\ref{tab:final-allH} along 
with their experimentally observed values and full assignment of all modes. 
The deviation of the vibrational frequencies calculated at different levels of 
theory with respect to the frequency obtained from the non-correlated 
VSCF/DF-SCS method are plotted in Figs.~\ref{fig:bar-p-1}a, \ref{fig:bar-p-1}b 
and \ref{fig:bar-p-1}c. The trend observed for the root mean squared deviation 
(RMSD) with respect to VSCF/DF-SCS going from harmonic (148\,cm$^{-1}$) to the 
inclusion of only diagonal anharmonic potentials (151\,cm$^{-1}$), and finally 
to the full potential expansion up to pair contributions at PM3 level 
(78\,cm$^{-1}$) and DF-MP2/cc-pVDZ level (24\,cm$^{-1}$) shows that  a balanced 
description of diagonal and coupling contributions is of importance in 
simplified models. Surprisingly, for VC-MP2 the RMSD (45\,cm$^{-1}$) 
deteriorates compared to the non-correlated VSCF treatment. Correlation 
correction may not be necessary for such a large system\cite{jung1996:10332}, 
and the VC-MP2 method may suffer from a degeneracy problem due to the high 
density of states in Methyl benzoate. 
As seen from Figs.~\ref{fig:bar-p-1}a, \ref{fig:bar-p-1}b and \ref{fig:bar-p-1}c, the VC-MP2 vibrational frequencies deteriorate for modes 2, 14, 16, 17, 35,
41, 45 and 46. The V-MP2 also shows a degeneracy problem for these modes. Other
than for these modes, VC-MP2 method shows a generally good agreement with the 
experimental results.  

For low frequency modes dual level calculations with PM3 pair couplings are 
not reliable.  The RMSD calculated for all  40 low frequency modes is thus quite
high, 73\,cm$^{-1}$. Harmonic and diagonal frequencies (frequency calculated from
the  anharmonic diagonal PES) are also quite off for several modes (1, 2, 3, 
13, 18, 35, 41 to 48) from the reference frequencies and give RMSDs of 
35\,cm$^{-1}$ and 37\,cm$^{-1}$, respectively. Frequencies calculated with the
non-correlated VSCF method with DF-MP2/cc-pVDZ diagonals and pair coupling
potentials  are in good agreement with the experimental results, which yields
a very low RMSD of 14\,cm$^{-1}$ for the 40 low frequency modes. Dual level 
calculations using DF-MP2/aug-cc-pVTZ
diagonal and DF-MP2/cc-pVDZ pair potentials, improve the results further, with
a  RMSD of 11\,cm$^{-1}$. 

In the low frequency region mode 32, which corresponds to the asymmetric
bending motion in the methyl group, is the most problematic mode, as seen in
Fig.~\ref{fig:bar-p-1}b. Harmonic and diagonal
frequencies for this mode are 130\,cm$^{-1}$ and 192\,cm$^{-1}$ too high,
respectively. The frequency calculated employing DF-MP2/cc-pVDZ pair
potentials is also too high ($\sim$\,80\,cm$^{-1}$) compared to the 
experimental result. This may indicate a strong coupling with the methyl 
group rotation or a missing pronounced triple coupling within the methyl group, 
neither of which are well represented in our current model.

\subsection{High frequency vibrational modes}
The high-frequency vibrational modes (41 to 48) are spectrally isolated from
all other modes of Methyl benzoate (see Table~\ref{tab:final-allH}). The 
deviation in frequency for these 8 modes calculated at different levels of 
theory w.r.t. the VSCF/SCS-VDZ method is plotted in Fig.~\ref{fig:bar-p-1}c. 
Harmonic and diagonal frequencies are too high. The RMSD w.r.t. the 
VSCF/DF-SCS results for the C--H stretch vibrational modes (41 to 48) alone 
is 319\,cm$^{-1}$ for the harmonic frequency, whereas it is 454\,cm$^{-1}$ for 
the diagonal frequencies. The large RMSDs for these eight modes from the 
harmonic and the diagonal frequency calculations are mostly due to the mode 42,
which shows an unexpected red shift to the frequency $\sim\,2550$\,cm$^{-1}$ 
(this is discussed in details later). However, when pair couplings are 
included, the RMSD is reduced dramatically. Dual level calculations, with  
DF-MP2/aug-cc-pVTZ diagonal potentials and  PM3 pair potentials (see 
Table~\ref{tab:final-allH}) yield a RMSD of 100\,cm$^{-1}$. In this case mode 
42 is also quite high but shows a red shift from the conventional C--H 
stretch vibrational frequency to a frequency of 2787\,cm$^{-1}$.  Dual level 
non-correlated VSCF computation with DF-MP2/aug-cc-pVTZ diagonal and 
DF-MP2/cc-pVDZ pair potentials are surprisingly accurate for the C--H 
stretch vibrational frequencies. The RMSD calculated for these eight modes
is 13\,cm$^{-1}$ only. 
\begin{figure}[!h]
\begin{center}
\includegraphics[width=0.8\linewidth,angle=0.0,clip]{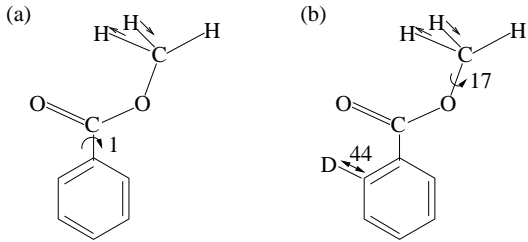}
\caption{(a) Vibrational motion of Methyl benzoate responsible for the 
unexpected red shift. (b) Secondary coupling of C--D and rotational methyl 
group.}\label{fig:wagging}
\end{center}
\end{figure}

A rather surprising result of  the anharmonic pair-coupling calculations is 
that one of the C--H stretch vibrational modes is shifted to a 
much lower frequency than the remaining C--H stretch vibrations. 
This red shift of the C--H stretch vibrational mode has not been discussed in
literature, so far. However, a close investigation of the experimental 
spectra\cite{sdbs},  reveals the presence of an unassigned peak at 
2560\,cm$^{-1}$, with low intensity. 
In a recent 1D IR experimental study of deuterated Methyl benzoate by Steinel
group\cite{steinel}, a low intensity peak is observed at 2542\,cm$^{-1}$ which
matches with the calculated frequency (see Fig.~\ref{fig:spectra-mb}). 

Closer inspection of this mode reveals that this frequency originates from 
the combination of anti- symmetric C--H (two out of plane hydrogens) stretch 
vibration in the methyl group along with the out of plane rotation of the 
ester group w.r.t. the phenyl ring about the C--C bond (mode 1). The ester 
group rotational motion pulls in one out-of-plane hydrogen to the plane of 
molecule and pushes the other from the molecular plane. Such a pull and push 
force changes the force constant of the anti- symmetric C--H stretch vibration, 
which unexpectedly lower the vibrational frequency of this band. Their motion 
is  depicted in Fig.~\ref{fig:wagging}a. The rotational motion is indicated 
by a curved arrow. Since this vibrational mode (mode 42) is, in the rectilinear
normal mode basis employed here, strongly coupled with the lower frequency 
mode 1, the harmonic and the diagonal anharmonic analysis fails to describe it,
where as pair coupling analysis explain it.

\subsection{C=O band}

The spectroscopically most studied vibrational band in peptides is the amide-I 
band. In the Methyl benzoate spectrum the C=O stretch vibrational band which is
analogous to amide-I band, is spectrally isolated from all other vibrational 
frequencies. Computationally, however, the study of this vibrational mode is 
not straightforward. The harmonic and the anharmonic diagonal frequencies of 
the C=O stretch vibration at DF-MP2/aug-cc-pVTZ level are relatively high 
(1766\,cm$^{-1}$ and 1759\,cm$^{-1}$ respectively) for  this mode,
compared to its experimental frequency (1724\,cm$^{-1}$) (see
Table~\ref{tab:final-allH}).  
Dual level frequency calculations employing DF-MP2/aug-cc-pVTZ diagonal
and the PM3 pair potential is even worse (see Table~\ref{tab:final-allH}). 
The calculated C=O frequency is improved when dual level calculation are
performed replacing the DF-MP2/cc-pVDZ diagonal potential for the C=O
vibrational mode alone by a DF-L-CCSD(T)/cc-pVTZ diagonal potential. A
perfect match with the experimentally observed frequency (1724\,cm$^{-1}$) is
obtained. For Methyl benzoate, the non correlated VSCF computation yields 
the C=O stretch vibrational frequency of 1726\,cm$^{-1}$, where as both 
V-MP2 and VC-MP2 yield 1727\,cm$^{-1}$. For the C=O stretch mode a high level 
description of the diagonal potential is thus of tantamount importance.

\section{Deuterated Methyl benzoate} 
\begin{figure}[!h]
\begin{center}
\includegraphics[width=1.0\linewidth,angle=0.0,clip]{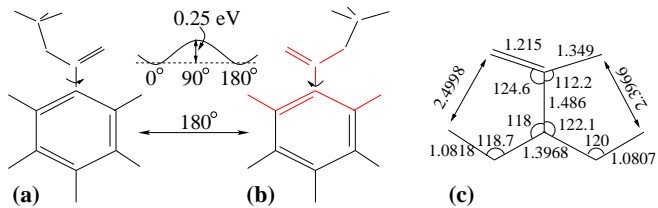}
\caption{Two structural conformers of Methyl benzoate and their approximate
  interconversion barrier height. (a) syn-o-deutero-methyl-benzoate (syn-o-DMB) (b) anti-o-deutero-methyl-benzoate (anti-o-DMB). (c) Possibility to form five member rings between ortho hydrogen of the phenyl ring and the ester group. Bond lengths are given in \AA~ and bond angles  in degree.}\label{fig:berrier-hight}
\end{center}
\end{figure}

Deuteration of Methyl benzoate in the ortho position of the benzene ring yields
the syn- and anti- isomers of ortho-deutero methyl benzoate (o-DMB)  shown in
Fig.~\ref{fig:berrier-hight}.   
In the potential energy minima the  ester group remains in the same plane as
the benzene ring. Upon thermal excitation, the conformers interconvert by a
rotation of the  ester group around  the C--C bond axis. 
The barrier height for the ester group rotation in Methyl benzoate is 
estimated at $\sim 0.25$\,eV, as 
calculated by the DF-MP2 method with aug-cc-pVTZ basis set. This is rather 
small and easily accessible even at room temperature and thus the conformers
can easily interconvert. Since both  conformers have nearly the same
equilibrium energy, they are also equally populated. Although both 
conformers have nearly the same molecular energy, they possess different 
vibrational C--D frequencies due to the different couplings to the ester 
group. With the best possible computation (dual level VSCF)  a C--D stretch 
frequency difference for these two conformers of about 10\,cm$^{-1}$ is found.

\subsection{Identification of isotope effects by co-diagonalization}

\begin{figure}[!h]
\vspace{-1.0cm}\includegraphics[width=\linewidth,angle=0.0,clip]{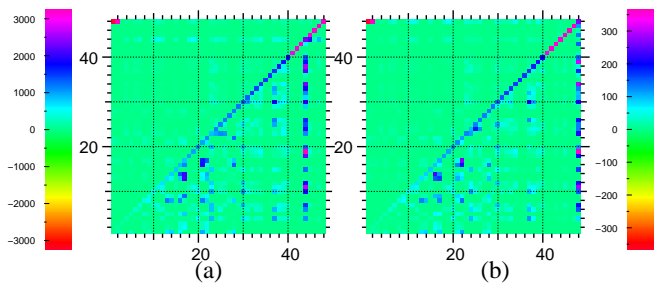}
\caption{Frequency spectra of Methyl benzoate isotopomers. 
(a) syn-o-deutero-methyl-benzoate  and (b)  anti-o-deutero-methyl-benzoate. 
See main test for explanation of the color scales and discussion of the 
depicted matrix elements.}\label{fig:freq-spectrum}
\end{figure}

The co-diagonalization\cite{roll1961:177} method finds a convenient application 
in vibrational spectroscopy to identify the coupling of different modes and 
isotope effects. When an atom is substituted with its isotope, some of the
eigen vectors differ from the non-substituted molecule. In co-diagonalization 
these are identified by non-zero off-diagonal elements. The frequency shift for
the primary isotope effect appears in the diagonal element and secondary
effects appear as off-diagonal elements. Fig.~\ref{fig:freq-spectrum} (a) and
(b) depict the resulting co-diagonalized matrices of the syn- and anti-o-DMB.
Fundamental frequencies are plotted on the diagonal and the residual couplings 
are shown in the upper left triangle on the same scale as the diagonal. The 
scale is given by the left rainbow color spectrum. Ten times magnified 
coupling elements are shown in the lower right triangle. The corresponding 
scale is given on the right side.

The syn- and anti-isomers show a clear primary isotope effect for the 44$^{th}$ 
(2171\,cm$^{-1}$) and 48$^{th}$ (2161\,cm$^{-1}$) vibrational modes, respectively.
The primary  isotope effect is observed in the C--D stretching vibration as 
expected. The secondary isotope effects are mainly due to 
the coupling with the C--D vibrational mode. These appear in the low frequency 
region. A noticeable feature of the isotope effect is that both isotopomers 
show secondary isotope effects for the same normal modes (modes 8, 13, 14, 17, 
18, 19, 20, 23, 27 and 31), where as primary isotope effects are  observed for 
different normal modes. A strong secondary isotope effect is observed at the 
upper left corner of Fig.~\ref{fig:freq-spectrum} (a) and (b), due to the 
coupling of C--D and methyl group rotation (see Fig.~\ref{fig:wagging}b).

\begin{figure}
\begin{center}
\includegraphics[width=1.0\linewidth,angle=0.0,clip]{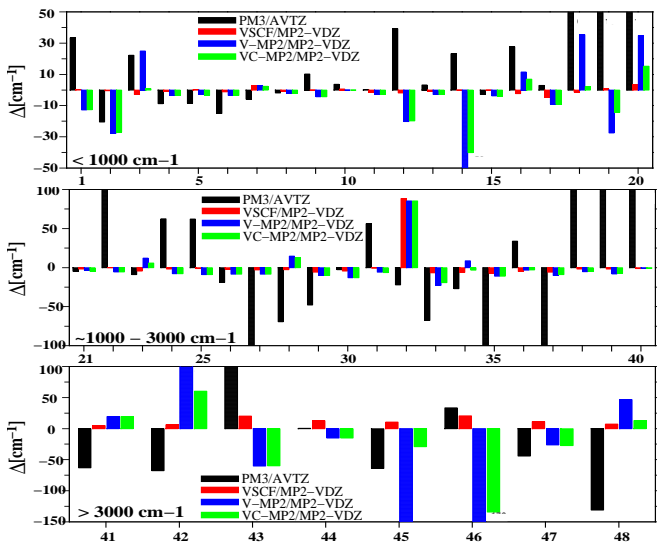}
\caption{A comparison of vibrational frequencies of anti-o-DMB calculated at
  different levels of theory w.r.t. VSCF/DF-SCS level of calculation.
 The  labels indicate the level of the theory was used for the calculations.}\label{fig:bar-p-d11-1}
\end{center}
\end{figure}

\subsection{Vibrational frequencies of anti-o-DMB}

For the deuterated species, the overall RMSD shows a similar behavior as in
undeuterated Methyl benzoate. Calculated harmonic and anharmonic diagonal 
frequencies are quite offset from the reference  frequencies and give high 
RMSD values of 137\,cm$^{-1}$ and 130\,cm$^{-1}$, respectively. Non-correlated 
VSCF dual level frequency calculations based on  MP2/aug-cc-pVTZ diagonal and 
PM3 pair potentials also deviate considerably from the reference frequencies 
(RMSD = 76\,cm$^{-1}$). 
The non-correlated VSCF frequencies employing DF-MP2/cc-pVDZ pair and
DF-MP2/aug-cc-pVTZ diagonal potentials show good agreement with
the reference frequencies and yield a very low RMSD of 14\,cm$^{-1}$, for all 48
modes.  Figs.~\ref{fig:bar-p-d11-1}a, \ref{fig:bar-p-d11-1}b and  \ref{fig:bar-p-d11-1}c depict
the general features of the different computational methods for the deuterated
Methyl benzoate and Table~\ref{tab:final-allH} presents the vibrational frequencies for the anti-o-deutero-methyl-benzoate.  Dual level frequency calculation 
employing DF-MP2/aug-cc-pVTZ
diagonal potentials and PM3 pair potentials for all 40 low frequency modes are
not reliable, but for the high frequency modes it yields quite reasonable
results. The harmonic and the diagonal anharmonic frequencies are also too
high in the low frequency region as well as the high frequency region. For the
low frequency modes (40 modes) the non-correlated VSCF employing
DF-MP2/aug-cc-pVTZ diagonal and  DF-MP2/cc-pVDZ pair potentials are in good
agreement with the reference frequencies, yielding a RMSD of 13\,cm$^{-1}$. This
level of computation is also adequate for the high frequency region and yields
a RMSD for the 8 C--H stretch vibrational modes of  17\,cm$^{-1}$.   

\subsection{C--D band}

\begin{figure*}
\begin{center}
\includegraphics[width=0.9\textwidth,angle=0.0,clip]{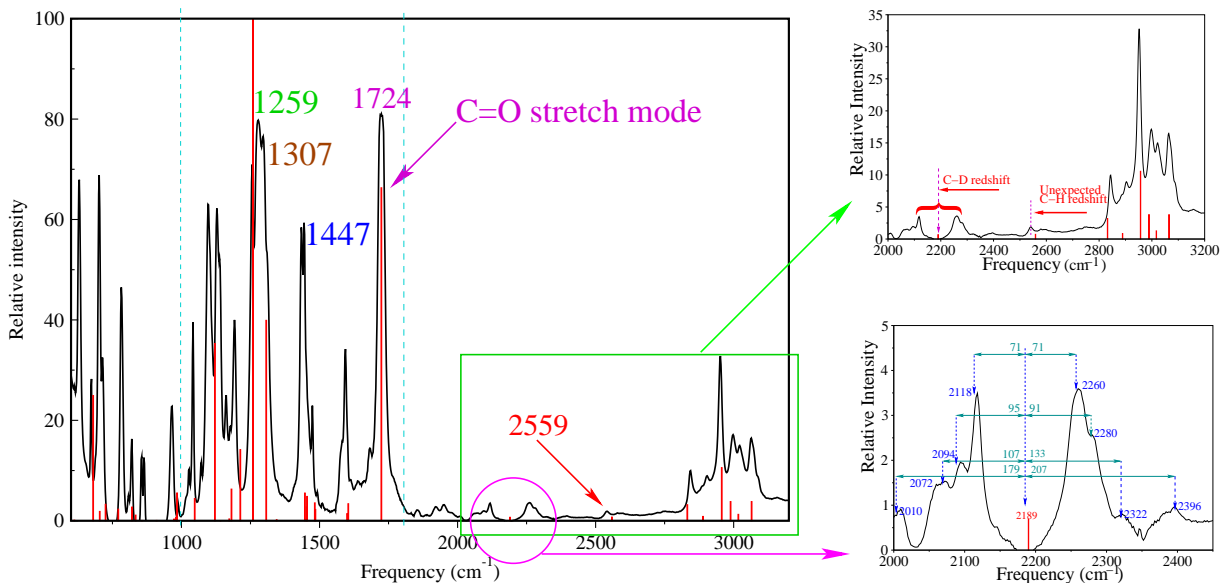}
\caption{1D IR experimental spectrum\cite{lemman} (black line) of deuterated Methyl benzoate is plotted along with calculated anharmonic vibrational frequencies (red sticks). Intensity of the calculated frequencies are from harmonic frequencies calculation.}\label{fig:spectra-mb}
\end{center}
\end{figure*}

The harmonic C--D vibrational frequencies at DF-MP2/aug-cc-pVTZ level are 
rather large, they are 2384\,cm$^{-1}$ and 2392\,cm$^{-1}$ for the syn- and 
anti-o-DMB, respectively. 
The anharmonic diagonal frequency calculation at DF-MP2/aug-cc-pVTZ level 
improved the results over harmonic calculation  by about 70\,cm$^{-1}$. 
It is noteworthy that the C--D stretching frequency is significantly improved 
by using the pair potentials calculated at DF-MP2/cc-pVDZ level.
The non-correlated dual level VSCF method yields C--D
vibrational frequencies of 2171\,cm$^{-1}$  and 2161\,cm$^{-1}$, for the two isotopomers. 

An interesting result is observed when the dual level calculations are
performed with systematically improved diagonal PES for the C--D stretch
vibrational mode. With the improved basis set (for all atoms same
basis sets are used) for the diagonal PES, the C--D anharmonic vibrational
frequency exhibits a red shift. The frequency change is rather large for the
DZ to TZ basis sets, and it converges quickly for larger basis sets. For
example, when the diagonal potential for C--D (in anti-o-DMB) vibrational 
mode is calculated with  augmented basis sets, a 32\,cm$^{-1}$ frequency
shift is observed going from AVDZ to AVTZ. On the other hand, a basis set
improvement from AVTZ to AVQZ shows just 1\,cm$^{-1}$ frequency 
shift (Table~\ref{tab:cd-d11} coloumn SCF). 
\begin{table}[!h]
\caption{The C--D vibrational frequency for anti-o-DMB.  The diagonal 
potential for the C--D stretch vibrational mode is calculated at different 
level of theory and employing different size of  basis sets. All other diagonal
and pair coupling potentials are calculated at DF-MP2/cc-pVDZ level. }
\label{tab:cd-d11}
\begin{ruledtabular}
\begin{tabular}{lllcccc}
isotope & Method & Basis & Diagonal & SCF & V-MP2 & VC-MP2\\
\hline
     & MP2     &DZ & 2367 & 2233 & 2243 & 2247 \\
syn  & MP2     &TZ & 2336 & 2187 & 2210 & 2210 \\
o-DMB& MP2    &ATZ & 2198 & 2180 & 2202 & 2201 \\
     & CCSD(T)& TZ & 2308 & 2161 & 2185 & 2184 \\
\hline
    & MP2     &DZ & 2374 & 2242 & 2257 & 2257 \\
    & MP2     &TZ & 2343 & 2196 & 2220 & 2219 \\
    & MP2     &QZ & 2338 & 2190 & 2214 & 2214 \\
    & MP2    &ADZ & 2359 & 2221 & 2239 & 2239 \\
anti& MP2    &ATZ & 2235 & 2189 & 2212 & 2212 \\
o-DMB& MP2   &AQZ & 2335 & 2188 & 2212 & 2211 \\
    & Ext    &AVXZ& 2331 & 2183 & 2208 & 2207 \\
    & CCSD(T)& TZ & 2315 & 2171 & 2194 & 2194 \\
    & CCSD(T)&T/QZ& 2310 & 2166 & 2190 & 2189 \\
    & PM3 & \multicolumn{2}{c}{ selected triple} & 2172 & 2197 & 2197\\
\end{tabular}
\end{ruledtabular}
\end{table}
The C--D shows also a red shift when
highly correlated methods are used. A non-correlated VSCF calculation where the
diagonal potential for the C--D vibrational mode is calculated at the
DF-L-CCSD(T)/cc-pVTZ level, yields C--D frequencies of 2161\,cm$^{-1}$ and 
2171\,cm$^{-1}$ for the syn- and anti-o-DMB respectively. When the DF-L-CCSD(T)
calculation is performed  using cc-pVQZ basis only for the
deuterium and the directly connected carbon (for all other atoms cc-pVTZ basis
are used), the frequency is lowered further by 5\,cm$^{-1}$ with respect to  the
DF-L-CCSD(T)/cc-pVTZ result (Table~\ref{tab:cd-d11}). 

A detailed analysis of the coupling pattern shows that only a few modes are 
coupled with the C--D stretch vibrational mode. It is observed that modes 17 
and 19 which correspond to the C--D bending modes, are strongly coupled with 
the C--D stretching mode. Also some other modes like 12, 13, 14, 15, 16, 18, 
21, 24,  have some weak coupling with the C--D stretch vibrational mode (see 
supplemental information). All these modes are involved with some kind of C--D
bending motions (see Table~\ref{tab:final-allH}).

One might expect that higher order terms in the many-body expansion of the
PES have a significant influence on the vibrational frequency of the anharmonic
system.  The PM3 triple coupling potentials were used for such an analysis. 
Selected PM3 level triple couplings were added to the DF-MP2/cc-pVDZ pair 
coupling. This lowers the C--D stretching frequency negligibly (see 
supplemental information) indicating a minor influence of triple 
couplings\cite{rauh2004:9313} for this particular mode.

There are two significant peaks observed in the linear IR absorption spectra 
(see lower right magnified region in Fig.~\ref{fig:spectra-mb}) which are 
equally separated from the calculated C--D stretch frequency at 2189\,cm$^{-1}$. 
It seems that the C--D vibrational stretching mode is coupled with some low 
vibrational modes and due to the Fermi resonance it split up into two bands. 
Vibrational CI calculation may be necessary to describe this feature.

\subsection{C=O band}
Due to the structural arrangement, there is a possibility to form a strained 
five membered ring between the ester and the phenyl ring, which may induce an 
additional coupling between the  C=O and the C--D band. The possible structure 
is shown in Fig.~\ref{fig:berrier-hight}c. Both the oxygen atoms in the ester 
group may take part in five membered rings along with the ortho and 
$\beta$-hydrogens of the phenyl ring. 
\begin{table}[!h]
\caption{Anharmonic vibrational frequency for the C=O stretch mode calculated
  at dual level VSCF method with DF-L-CCSD(T)/cc-pVTZ diagonal and
  DF-MP2/cc-pVDZ pair potentials.}
\label{tab:co-band}
\begin{ruledtabular}
\begin{tabular}{ccccc}
Isotope & SCF & V-MP2 & VC-MP2 & Experimental\\
\hline
syn-o-DMB  & 1724.5 & 1724.4 & 1724.5 &\\
anti-o-DMB & 1724.3 & 1724.1 & 1724.1 & 1724 \\
MB   & 1725.6 & 1727.2 & 1727.2 & \\
\end{tabular}
\end{ruledtabular}
\end{table}
The calculated possible non-bounded distances are  $2.499$\,{\AA} and
$2.397$\,{\AA} for the syn- and anti- positions, respectively, which are much
larger than the required bond length to form a true five membered
ring. Therefore, isotopic substitution at the ortho position in the phenyl
ring does not have any influence on the C=O vibrational frequency. The C=O
vibrational frequency remains largely unchanged for both syn- and anti-o-DMB. 
Calculated anharmonic frequencies in dual level calculations yield the same
frequency of 1724\,cm$^{-1}$ for both syn- and anti- conformers, which is in
agreements with the experimental results. 

\subsection{Anharmonicity observed in the VSCF calculations}

\begin{table}[!h]
\caption{Diagonal anharmonicity for vibrational frequencies of
  Methyl benzoate and its two isotopomers are calculated at the 
non-correlated VSCF method.  } 
\label{tab:diag-anharmoni}
\begin{ruledtabular}
\begin{tabular}{crrrcrrr}
Mode&\multicolumn{3}{c}{Isotopomer} & Mode&\multicolumn{3}{c}{Isotopomer}\\
\cline{2-4}\cline{6-8}
No. & MB & syn- & anti- & No. & MB & syn- & anti- \\
\hline
1  &  17.67  &  18.98  &  17.57  &  21 &  26.03  &  29.62  &  24.43  \\
2  &  17.45  &  16.79  &  16.67  &  22 &   5.73  &   1.67  &   3.41  \\
3  &  83.80  &  78.25  &  79.38  &  23 &  21.06  &  20.36  &  21.55  \\
4  &   4.16  &   4.14  &   3.83  &  24 &  -2.10  &   0.10  &   4.42  \\
5  &   9.05  &   8.95  &   8.80  &  25 &  13.86  &  13.19  &  12.17  \\
6  &   3.78  &   3.77  &   3.73  &  26 &   9.72  &   5.72  &   1.34  \\
7  &  -5.71  &  -2.53  &  -2.24  &  27 &  10.37  &  10.25  &   9.97  \\
8  &   3.66  &   4.42  &   3.56  &  28 &  22.96  &  22.88  &  19.70  \\
9  &   5.20  &   5.81  &   5.30  &  29 &   5.65  &   6.16  &   5.88  \\
10 &   0.29  &   0.73  &   0.79  &  30 &  -3.19  &  -8.34  &  -6.33  \\
11 &   1.94  &   1.92  &   1.92  &  31 &   4.96  &   7.17  &   4.45  \\
12 &  -0.47  &  -0.92  &  -1.31  &  32 &   1.51  &  10.88  &   9.98  \\
13 &  21.11  &  21.71  &  19.16  &  33 &   5.02  &   4.20  &   6.72  \\
14 &  15.04  &  14.94  &  15.46  &  34 &   3.81  &   6.24  &   5.20  \\
15 &   1.32  &  -0.12  &   0.30  &  35 &   1.17  &   0.93  &   1.22  \\
16 &   7.82  &   7.00  &   5.51  &  36 & -10.23  & -10.05  & -10.13  \\
17 &  25.71  &   4.03  &   3.69  &  37 &   2.92  &   3.39  &   2.37  \\
18 &   1.58  &   9.65  &   9.94  &  38 &  -2.05  &  -1.32  &  -1.85  \\
19 &  11.41  &  12.56  &  15.04  &  39 &   0.04  &   0.93  &   0.26  \\
20 &   3.32  &  -5.11  &  -5.94  &  40 &  -6.85  &  -7.09  &  -6.60  \\
\end{tabular}
\begin{tabular}{crrr}
Mode&\multicolumn{3}{c}{Isotopomer}\\
\cline{2-4}
No. & MB & syn- & anti- \\
\hline
41 & -120.76 & -28.66  & -29.07  \\
42 & -278.75 & -214.25 & -216.96 \\
43 & -155.03 & -123.61 & -123.67 \\
44 & -198.70 & -170.12 & -124.85 \\
45 & -88.52  & -114.61 & -68.48  \\
46 & -102.29 & -90.28  & -87.35  \\
47 & -88.28  & -87.37  & -85.05  \\
48 & -67.71  & -70.31  & -94.47  \\
\end{tabular}
\end{ruledtabular}

\end{table}
The diagonal anharmonicities in frequency calculations are presented in
Table~\ref{tab:diag-anharmoni} where mode numbers are based on the harmonic
normal mode analysis. The diagonal anharmonicities are very high for the high
frequency C--H stretch vibrational modes, up to a few hundred
wavenumbers. Especially  mode 42, which is unexpectedly red shifted, shows 
maximum anharmonicity. Other than the C--H stretch vibrational modes, diagonal
anharmonicities are rather small ($<10$\,cm$^{-1}$). Few low frequency modes 
show anharmonicities slightly higher than 10\,cm$^{-1}$. These vibrational 
modes are mostly involved with  C--H bending motions 
(see Table~\ref{tab:final-allH}).
For the first three modes the calculated vibrational
frequencies are not reliable and for that reason anharmonicities are also
unreliable.  

\subsection{Off-diagonal anharmonicity}
Off-diagonal anharmonicities calculated  for the few spectroscopically most 
important modes are presented in Table.~\ref{tab:offdiag-anharmoni}. 
Calculated off-diagonal anharmonicities for the C=O vs. C--D coupling modes 
for both the isotopomers are less than a wave number.  
\begin{table}[!h]
\caption{Off diagonal anharmonicities for the C=O and C--D coupling modes
  calculated with non-correlated VSCF method with DF-MP2/AVTZ diagonal and
  DF-MP2/VDZ pair potentials. } \label{tab:offdiag-anharmoni} 
\begin{ruledtabular}
\begin{tabular}{clcccc}
Coupling & level & $\omega_a$ & $\omega_b$ & $\omega'$ & $\Delta\omega$ \\
\hline
& SCF & 1771.03 & 2233.49 & 4003.846 & 0.67 \\
CO CD$_7$ & V-MP2 & 1768.81 & 2242.75 & 4012.076 & -0.52 \\
& VC-MP2 & 1768.81 & 2246.59 & 4015.142 & 0.36 \\
\hline
& SCF & 1770.62 & 2242.33 & 4012.771 & 0.18 \\
CO CD$_{11}$ & V-MP2 & 1768.75 & 2257.22 & 4025.613 & 0.36 \\
& VC-MP2 & 1768.75 & 2257.22 & 4025.654 & 0.32 \\
\end{tabular}
\end{ruledtabular}
\end{table}
Such a negligibly small off-diagonal anharmonicity also indicates that there
is negligible direct coupling between the C=O and the C--D vibrational modes. 

\section{Conclusions}
It has been observed that harmonic frequencies are  very poor approximation to
assign the vibrational frequencies of Methyl benzoat and its two isotopomers. 
Anharmonic diagonal frequencies show some improvements but were still not 
sufficient to reach a reasonable assignment. The fundamental transition 
frequencies calculated with the VSCF method from a pair potential energy 
surface expansion seem promising, especially when the 2D PES are calculated at 
a sufficiently high ab initio level. The success of the VSCF frequency 
calculations depends upon the accuracy of the PES, in particular near the 
equilibrium where denser grid points are required. Dual level computations in 
which the diagonal anharmonic potential along a single vibrational mode is 
calculated using higher level ab initio methods than for coupling potentials 
provide an efficient route to the computation of the PES expansion in the VSCF 
framework. Such a dual level VSCF calculation with DF-L-CCSD(T)/cc-pVTZ 
diagonal and DF-MP2/cc-pVDZ pair coupling potentials provided a nearly perfect 
agreement of the C=O vibrational frequency with respect to the experimental 
result. 

Carrying out a systematic study, we have shown that not all pair couplings are
necessary to describe a particular vibrational band. Using only 14 coupling
potentials for the C--D stretch vibrational mode the computed frequency is in
good agreement with the result based on the full set of 1128 couplings.
Non-uniform IMLS interpolation has been successfully used to reduce the  
computational cost for potential energy surface generation even further. 

An unexpected red shift has been observed for a C--H stretch vibrational mode
when the VSCF calculation has been performed with pair coupling
potentials. The inability to find this mode in harmonic and diagonal 
anharmonic calculations indicates that this is a concerted anharmonic effect 
and pair and higher order couplings are in fact necessary to understand this
feature. Our investigations for the C--H(D) stretch vibrational modes with
selected triple couplings at PM3 level do not improve the results much over
the ab initio pair coupling calculations. This could indicate that either PM3
triple coupling potentials are in sufficient or that higher order coupling
effects are negligible in this model.

The correct assignment of the C--D stretch frequency still poses a problem
to both the theoretician and experimentalist. The calculated C--D stretch
frequency just sits between the two strong peaks observed in the linear IR
absorption spectra at the expected C--D frequency region. It seems that the
C--D stretch frequency mode is coupled with some low vibrational frequency 
modes and thus it is  shifted equally in both directions and appears as two 
peaks. Vibrational CI calculation may be necessary to describe this feature. 

The diagonal and the off-diagonal anharmonicities have been calculated by
non-correlated VSCF method. Other than for a few low frequency modes, which are
involved in the C--H bending modes, diagonal anharmonicities are
very small. The negligible off-diagonal anharmonicity for C=O and C--D 
coupling modes indicates that these modes are mostly decoupled.

\begin{acknowledgments}
This work has been supported by Deutsche Forschungsgemeinschaft. Computational facility
from Leibniz-Rechenzentrum is gratefully acknowledged.
\end{acknowledgments}

%


\end{document}


\preprint{AIP/123-QED}

\title{Vibrational Spectroscopy of Methyl benzoate}

\author{Kiran Sankar Maiti}
 \email{kiran.maiti@ch.tum.de.}
 \affiliation{Department of Chemistry and Molecular Biology, University of Gothenburg, Box 462, SE-40530 Gothenburg, Sweden.}
\author{Christoph Scheurer}%
\affiliation{ 
Department of Chemistry, 
Technical University of Munich, D-85747 Garching, Germany. 
}%

\date{\today}

\maketitle

\section{Supplementary information}
\begin{table}[!h]
\caption{Strongly coupled modes with C--D vibrational mode for anti-o-deutero-methyl-benzoate. Although the frequency differ a little  amount, but the qualitative result does not depend upon the isotopomers.}
 \label{tab:coup-d11}
\begin{ruledtabular}
\begin{tabular}{lrrrr}
\multicolumn{1}{c}{Coupled modes}& SCF & V-MP2 & VC-MP2 \\
\hline
all decoupled & 2316.4 & 2316.4 & 2316.4 \\
17  & 2287.4 & 2295.9 & 2295.9 \\
19  & 2294.8 & 2294.7 & 2294.7 \\
17,19 & 2265.6 & 2274.7 & 2274.5 \\
12,17,19 & 2243.1 & 2258.9 & 2258.7 \\
12,14,17,19 &  2232.3 & 2249.8 & 2249.5 \\
12,14,15,17,19 & 2226.4 & 2244.3 & 2243.9\\
12,14,15,17,19,21 & 2219.9 & 2238.8 & 2238.5 \\
12,14,15,17,19,21,24 & 2215.6 & 2234.1 & 2233.8 \\
12,13,14,15,17,19,21,24 & 2211.4 & 2230.3 & 2229.9\\
12,13,14,15,16,17,19,21,24 & 2204.5 & 2224.8 & 2224.4\\
12,13,14,15,16,17,18,19,21,24 & 2199.2 & 2220.2 & 2219.8\\
8,12,13,14,15,16,17,18,19,21,24 & 2192.3 & 2214.1 & 2213.6\\
7,8,12,13,14,15,16,17,18,19,21,24 & 2187.9 & 2209.9 & 2209.5\\
7,8,12,13,14,15,16,17,18,19,21,24,30 & 2183.8 & 2205.9 & 2205.5\\
7,8,9,12,13,14,15,16,17,18,19,21,24,30 & 2180.1 & 2202.7 & 2202.2\\
\end{tabular}
\end{ruledtabular}
\end{table}

\begin{table}[!h]
\caption{Strongly coupled PM3 triple contributions for C--D vibrational mode
  of anti-o-deutero-methyl-benzoate. Although the frequency differ a little  amount, but the qualitative
  result does not depend upon the isotopomers.} 
\label{tab:triple-coup-d11}
\begin{ruledtabular}
\begin{tabular}{rlrrrr}
\multicolumn{2}{c}{Triple coupled modes}& SCF & V-MP2 & VC-MP2 \\
\hline
\multicolumn{2}{c}{No triples}  & 2003.2 & 2038.9 & 2038.3 \\
17,j,40 & 19 & 2167.9 & 2192.1 & 2191.5 \\
17,j,40 & 12,19 & 2168.4 & 2192.4 & 2191.8 \\
17,j,40 & 12,14,19 & 2168.0 & 2192.0 & 2191.4 \\
17,j,40 & 12,14,15,19 & 2172.6 & 2196.7 & 2196.1 \\
17,j,40 & 12,14,15,19,21 & 2172.2 & 2196.2 & 2195.6 \\
17,j,40 & 12,14,15,19,21,24 & 2173.9 & 2198.9 & 2198.3 \\
17,j,40 & 12,13,14,15,19,21,24 & 2173.1 & 2198.6 & 2198.0 \\
17,j,40 & 12,13,14,15,16,19,21,24 & 2172.5 & 2198.0 & 2197.4 \\
17,j,40 & 12,13,14,15,16,18,19,21,24 & 2171.6 & 2197.2 & 2196.6 \\
\end{tabular}
\end{ruledtabular}
\end{table}